\documentclass[letter]{aa}

\usepackage{graphicx}
\usepackage{float}
\usepackage{placeins}
\usepackage{amsmath}
\usepackage{txfonts}
\usepackage{natbib}
\bibpunct{(}{)}{;}{a}{}{,}
\usepackage{xcolor}
\usepackage[hidelinks]{hyperref}
\usepackage{orcidlink}

\newcommand{\wsi}{[WSI2008]~778}
\newcommand{\lh}{MGSD~LH~117-43A/140}
\newcommand{\sk}{Sk~$-69$~135}
\newcommand{\kms}{\ensuremath{\mathrm{km\,s^{-1}}}}
\newcommand{\niv}{\ensuremath{\mathrm{N\,IV}}}
\newcommand{\niii}{\ensuremath{\mathrm{N\,III}}}
\newcommand{\nv}{\ensuremath{\mathrm{N\,V}}}
\newcommand{\hei}{\ensuremath{\mathrm{He\,I}}}
\newcommand{\heii}{\ensuremath{\mathrm{He\,II}}}

\begin{document}

\title{Massive stars in the SDSS-V survey:\\
New O2 stars in the Large Magellanic Cloud}

\titlerunning{New O2 stars in the LMC}

\author{
A. Roman-Lopes\orcidlink{0000-0002-1379-4204}\inst{1}
\and
J. G. Fernández-Trincado\orcidlink{0000-0003-3526-5052}\inst{4}
\and
Marina Kounkel\orcidlink{0000-0002-5365-1267}\inst{5}
\and
A. A. C. Sander\orcidlink{0000-0002-2090-9751}\inst{6}
\and
Carlos G. Román-Zúñiga\orcidlink{0000-0001-8600-4798}\inst{2}
\and
A. Z. Lugo-Aranda\orcidlink{0000-0001-9226-9178}\inst{3}
\and
Sergio Sánchez-Sanjuán\orcidlink{0000-0002-2269-9348}\inst{2}
\and
Abhinna Sundar Samantaray\orcidlink{0000-0002-0635-2264}\inst{6}
\and
José Eduardo Méndez Delgado\orcidlink{0000-0002-6972-6411}\inst{7}
\and
Christian Moni Bidin\orcidlink{0009-0004-3783-6378}\inst{8}
}

\authorrunning{Roman-Lopes et al.}

\institute{
Departamento de Astronomía, Universidad de La Serena,
Av. Raúl Bitrán 1305, La Serena 1700000, Coquimbo, Chile
\email{aroman@userena.cl}
\and
Instituto de Astronomía, Universidad Nacional Autónoma de México,
A.P. 106, Ensenada 22800, Baja California, Mexico
\and
Universidad Andrés Bello, Facultad de Ciencias Exactas,
Departamento de Física y Astronomía -- Instituto de Astrofísica,
Autopista Concepción--Talcahuano 7100, Talcahuano, Chile
\and
Centro de Investigación en Astronomía,
Facultad de Ingeniería, Ciencia y Tecnología,
Universidad Bernardo O'Higgins,
Av. Viel 1497, Santiago 8370993, Chile
\and
Department of Physics and Astronomy,
University of North Florida,
1 UNF Dr, Jacksonville, FL 32224, USA
\and
Zentrum für Astronomie der Universität Heidelberg,
Astronomisches Rechen-Institut,
Mönchhofstr. 12--14, 69120 Heidelberg, Germany
\and
Instituto de Astronomía,
Universidad Nacional Autónoma de México,
A.P. 70-264, 04510 Mexico City, Mexico
\and
Instituto de Astronom\'ia, Universidad Cat\'olica del Norte,
Av. Angamos 0610, Antofagasta, Chile
\email{cmoni@ucn.cl}
}

\date{Received ...; accepted ...}

\abstract
{O2 stars define the hottest end of the normal O-star sequence, but their
identification requires blue-optical nitrogen diagnostics absent from many
surveys.}
{We reassess three luminous Large Magellanic Cloud (LMC) sources with multi-epoch
Sloan Digital Sky Survey V (SDSS-V) spectroscopy obtained with the Baryon
Oscillation Spectroscopic Survey (BOSS) spectrographs: \wsi, \lh, and \sk.}
{We measure visit-resolved radial velocities and H, He, and N equivalent widths
(EWs), compare robust combinations with LMC O2 templates at a common effective
resolution, and use OSTAR2002 spectral energy distribution (SED) fits and a Gaia
colour--magnitude diagram (CMD) as broad-band checks.}
{The weighted
$\log|{\rm EW}(\niv~\lambda4058)/{\rm EW}(\niii~\lambda4640)|$ values are
$0.64\pm0.29$, $0.48\pm0.23$, and $0.34\pm0.44$ for \wsi, \lh, and \sk.
The detection of \nv~$\lambda\lambda4604,4620$ supports the O2~If$^\ast$ and O2~V--III
classifications adopted for \wsi\ and \lh, respectively. For \sk, the uncertain \niv/\niii\ ratio and non-detection of
\nv\ motivate O2--3~V--III. Weak \hei~$\lambda4471$ may indicate unresolved
later-type contamination in \wsi\ and \sk. The radial velocities of \sk\ show
suggestive but non-significant variability. OSTAR2002 fits over 45--55~kK and
the Gaia CMD place all three sources in the luminous very-early-O regime;
\wsi\ shows a strong mid-infrared (MIR) excess, \lh\ is nearly photospheric,
and \sk\ shows only a modest $3$--$8\,\mu$m excess.}
{The BOSS spectra add \wsi\ and \lh\ to the LMC very-early-O population
and identify \sk\ as an additional O2--3 source. These cases show how incomplete
blue coverage and ambiguous catalogue associations can conceal the hottest
massive stars.}


\maketitle
\nolinenumbers

\section{Introduction}
\label{sec:introduction}

O2 stars define the hottest end of the normal O-star sequence and are rare
benchmarks for hot-star atmosphere, wind, and evolution models
\citep{Walborn2002,CrowtherWalborn2011,Martins2002,Kudritzki2002}. Their
blue-optical classification is set chiefly by the
\niv~$\lambda4058$/\niii~$\lambda\lambda4634$--4642 balance: \niv\ strongly
dominates \niii, \hei~$\lambda4471$ is absent or very weak, and detected
\nv~$\lambda\lambda4604,4620$ further supports the extreme ionisation;
\heii~$\lambda4686$ and the Balmer lines primarily constrain wind morphology
and luminosity class \citep{Walborn2002,RiveroGonzalez2012,CrowtherWalborn2011}.
Missing blue coverage, blending, and nebular emission can therefore produce
substantially later classifications.
\par

We reassess three Large Magellanic Cloud (LMC) sources with multi-epoch Sloan
Digital Sky Survey V (SDSS-V) spectroscopy obtained with the Baryon Oscillation
Spectroscopic Survey (BOSS) spectrographs. \wsi\ was previously linked to young
stellar object (YSO) and post-asymptotic giant branch (post-AGB) selections
\citep{Whitney2008,vanAarle2011}; \lh\ was classified as O3~III(f$^\ast$) and
later as O5~V based on Multi Unit Spectroscopic Explorer (MUSE) data lacking the
decisive blue nitrogen lines
\citep{GarmanyWalborn1987,Massey1989LH117,McLeod2019N44N180}; and \sk\ was
catalogued only as an OB source \citep{Sanduleak1970}. Our analysis combines
visit-resolved spectroscopy, LMC O2 template comparisons, and broad-band
checks.
\par

\begin{figure*}[!t]
\centering
\includegraphics[width=0.77\textwidth]{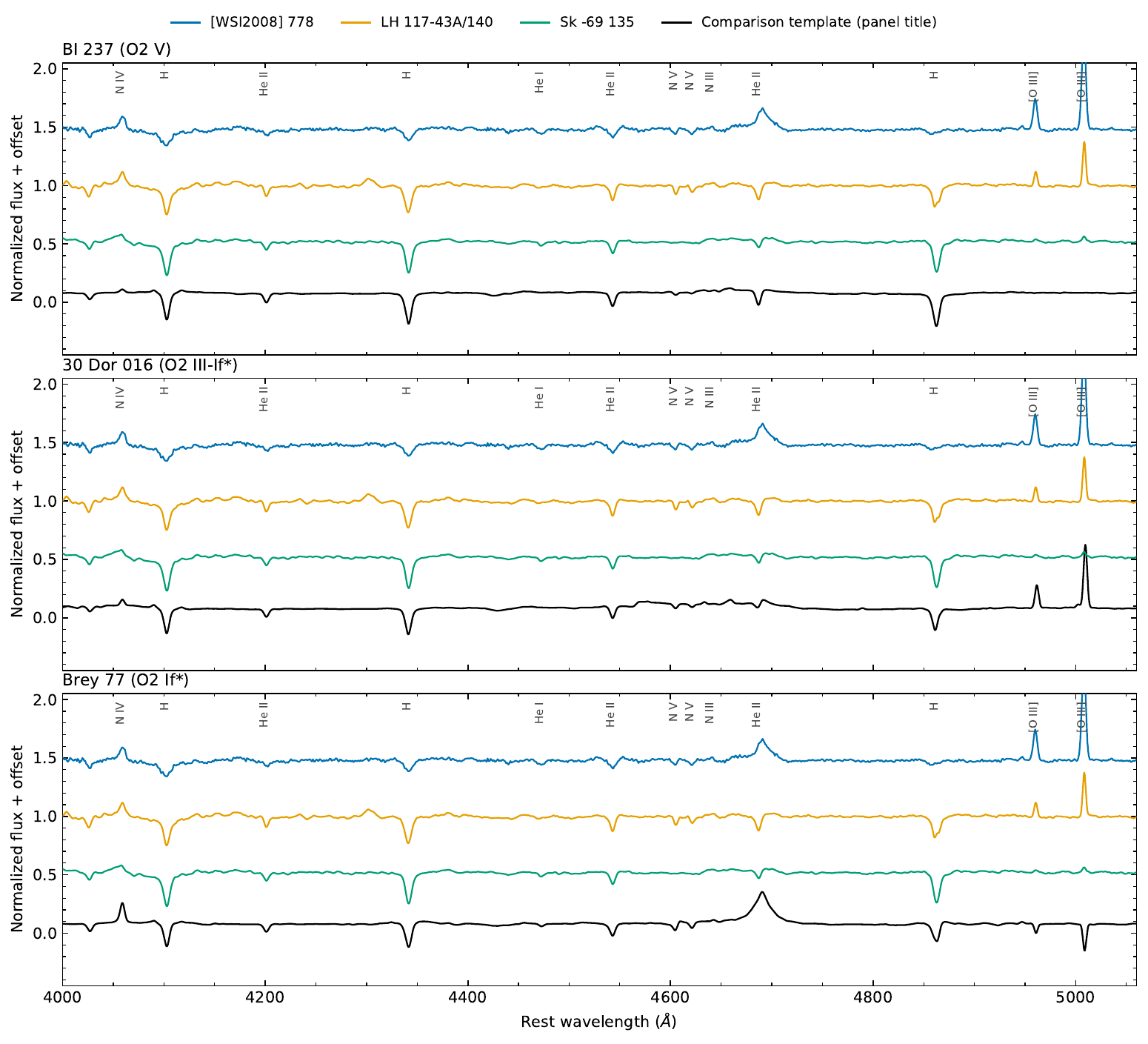}
\caption{Rest-frame normalised spectra of the three targets and LMC O2 templates. For this morphological comparison only, all spectra were matched to the same wavelength-dependent effective resolution defined by the largest BOSS \texttt{WRESL} among the science spectra. Narrow nebular [O~{\sc iii}]~$\lambda\lambda4959,5007$ emission in \wsi\ and \lh\ is accompanied by partial Balmer-line infilling.}
\label{fig:three_stars_o2_templates}
\end{figure*}

\section{Observations and analysis}
\label{sec:observations}

The southern SDSS-V Milky Way Mapper (MWM) spectra obtained with BOSS cover the
blue classification region and extend beyond H$\alpha$
\citep{BowenVaughan1973,Smee2013,Kollmeier2026}. Four visits are available for
\wsi\ and three each for \lh\ and \sk. Each visit was rest-frame corrected
with photospheric He~II, locally normalised, and measured independently;
equivalent widths (EWs) were combined using inverse-variance weights, with positive
values denoting absorption. Robust visit combinations were used only for
display, while quantitative measurements remained visit-resolved
(Appendix~\ref{app:boss}).
\par

The effective resolution was measured directly from the BOSS \texttt{WRESL}
arrays rather than from the nominal instrumental range. Across the three
sources, $R$ spans about 990--1374 at 4000~\AA\ and 1322--1757 at 5060~\AA.
For Fig.~\ref{fig:three_stars_o2_templates}, the science spectra and BI~237
(O2~V), 30~Dor~016 (O2~III--If$^\ast$), and Brey~77 (O2~If$^\ast$) were
matched to the same wavelength-dependent effective resolution; the EW and
velocity measurements use the original spectra. Source-specific values and the
convolution procedure are given in Appendix~\ref{app:templates}.
\par

\section{Spectral classifications}
\label{sec:classification}

All three targets show the high-ionisation morphology of the earliest O stars:
\niv~$\lambda4058$ is in emission, \niii~$\lambda\lambda4634$--4642 is weak,
and He~II absorption is prominent (Fig.~\ref{fig:three_stars_o2_templates}).
The weighted
$\log|{\rm EW}(\niv~\lambda4058)/{\rm EW}(\niii~\lambda4640)|$ values are
$0.64\pm0.29$, $0.48\pm0.23$, and $0.34\pm0.44$ for \wsi, \lh, and \sk;
\nv~$\lambda\lambda4604,4620$ is detected in the first two but not
significantly in \sk. Full EWs and statistical caveats are given in
Table~\ref{tab:key_lines_three} and Appendix~\ref{app:nitrogen}. Narrow
nebular [O~{\sc iii}] in \wsi\ and \lh\ indicates partial H$\beta$ infilling
and possible H$\gamma$ core contamination, so these Balmer cores are not used
as purely stellar diagnostics; \heii~$\lambda4686$ provides the cleaner wind
comparison.
\par

For \wsi, the strong \niv/\niii\ imbalance and detected \nv\ establish O2.
Its strong \heii~$\lambda4686$ emission places it in the If$^\ast$ morphology,
so we adopt O2~If$^\ast$. Reproducible \hei~$\lambda4471$
($0.31\pm0.10$~\AA) is stronger than expected for an isolated O2 spectrum and
may indicate an unresolved later-type contribution, which cannot be decomposed
at BOSS resolution.
\par

For \lh, \niv\ strongly dominates weak \niii\ and both \nv\ components are
detected, securely supporting O2. \heii~$\lambda4686$ remains in absorption;
however, the modest BOSS resolving power ($R\simeq1170$--1260 across the main
blue diagnostics) and nebular H$\beta$ infilling prevent a robust V/III
distinction. We therefore adopt O2~V--III. This also explains the previous O5~V classification based on MUSE data, whose
wavelength coverage begins near 4750~\AA\ and excludes the defining blue N
sequence.
\par

For \sk, \niv\ emission and weak \niii\ establish the very-early-O regime and
the absorption-dominated spectrum remains close to BI~237. The uncertain
\niv/\niii\ ratio and non-detection of \nv\ nevertheless motivate the more
conservative O2--3~V--III classification. Weak \hei~$\lambda4471$
($0.22\pm0.09$~\AA) is reproducible and may likewise reflect a later-type
contribution.
\par

\section{Kinematics and physical context}
\label{sec:context}

The weighted stellar velocities are $295\pm17$, $240\pm7$, and $277\pm7~\kms$
for \wsi, \lh, and \sk. Local nebular [O~{\sc iii}] gives
$280.8\pm4.3$, $251.3\pm4.4$, and $281\pm19~\kms$, respectively, so none of
the stellar-nebular differences is significant. For \sk, three visits over
23 days show a first-to-last change of $-39.2\pm17.8~\kms$; a constant-velocity
fit gives $\chi^2=5.33$ for two degrees of freedom ($p=0.069$), suggestive but
insufficient to establish binarity.
\par

The targets are not concentrated around 30~Doradus: their projected separations
from R136 are about 0.66, 1.14, and 0.93~kpc for \wsi, \lh, and \sk. In
particular, \lh\ lies in the LH~117/NGC~2122 association within the N180
H\,{\sc ii} complex \citep{Massey1989LH117,McLeod2019N44N180}. Thus they sample
distinct massive-star environments across the LMC, consistent also with the
agreement between stellar and local nebular velocities.
\par

\begin{figure}[!t]
\centering
\includegraphics[width=0.76\columnwidth]{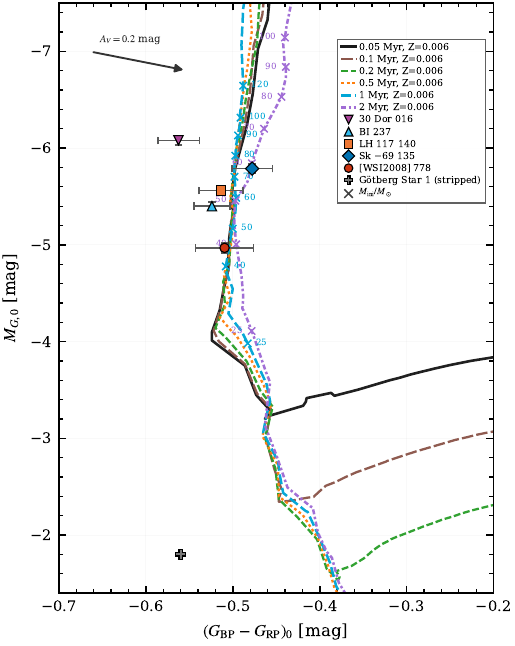}
\caption{Dereddened Gaia CMD for the three targets and BI~237/30~Dor~016. PARSEC
v1.2S $Z=0.006$ isochrones span 0.05--2~Myr; the arrow gives the displacement
for $A_V=0.2$~mag. G\"otberg Star~1 is the optically brightest stripped-star
comparison; O-star error bars show the 16th--84th percentile Monte-Carlo
intervals.}
\label{fig:gaia_cmd}
\end{figure}

As an independent broad-band check, we fit selected optical--near-infrared photometry
with half-solar OSTAR2002 atmospheres. Because these bands sample the
Rayleigh--Jeans side of such hot spectral energy distributions (SEDs), the
photometry does not determine a
unique $T_{\rm eff}$; instead, the spectroscopically motivated 45--55~kK grid
is sampled with equal weight (Appendix~\ref{app:sed}). The resulting positions in the Gaia colour--magnitude diagram (CMD),
$(G_{\rm BP}-G_{\rm RP})_0,M_{G,0}$, are $(-0.509,-4.97)$,
$(-0.514,-5.56)$, and $(-0.478,-5.79)$ for \wsi, \lh, and \sk, respectively,
compared with
$(-0.524,-5.40)$ for BI~237 and $(-0.562,-6.08)$ for 30~Dor~016. All three targets therefore lie in the same luminous blue part of the CMD as
established very-early-O stars. Figure~\ref{fig:gaia_cmd} shows PAdova and TRieste Stellar Evolution
Code (PARSEC) v1.2S $Z=0.006$ isochrones from
0.05 to 2~Myr and an $A_V=0.2$ reddening vector; these curves are shown only as a guide, not to provide precise ages or masses.
\par

Hot stripped helium stars can overlap very-early-O stars in intrinsic colour,
but the observed Magellanic-Cloud sample is much fainter optically
\citep{Gotberg2018,Gotberg2023}. Stars~1--8 span
$T_{\rm eff}\simeq57$--90~kK and $\log L/L_\odot\simeq3.5$--5.1; however,
their optical absolute magnitudes lie in the range
$-1.7\lesssim M_{V,\rm AB}\lesssim+0.8$. The optically brightest, Star~1, is
shown in Fig.~\ref{fig:gaia_cmd}. The science targets lie several magnitudes
above this regime, so luminosity rather than intrinsic colour is the main
discriminant. Their CMD positions strongly disfavour an ordinary
intermediate-mass stripped star as the optically dominant component while not
excluding past binary interaction, unresolved companions, or more massive
binary products.
\par

The OSTAR2002 scaling independently points to the same luminosity regime. The
numerical radius/extinction posteriors are reported in Appendix~\ref{app:sed};
qualitatively, \wsi\ shows a strong rising mid-infrared (MIR) excess, \lh\ remains nearly
photospheric, and \sk\ shows only a modest $3$--$8\,\mu$m excess.
\par

\section{Discussion}
\label{sec:discussion}

The spectroscopic and broad-band results constrain different aspects of the
problem. The adopted subtypes are morphological, based on the N~IV--N~III
balance, \nv\ where measurable, and the wind-sensitive He~II/Balmer
morphology; the 45--55~kK SED interval is a physically motivated nuisance range
for luminosity scaling, not an independent photometric $T_{\rm eff}$
measurement. This distinction matters at the earliest O subtypes, where the
mapping between morphology and $T_{\rm eff}$ also depends on gravity, wind
density, metallicity, and abundance pattern.
\par

Although all three sources show similarly high ionisation, their wind morphologies
differ. \wsi\ combines O2 nitrogen diagnostics with strong
\heii~$\lambda4686$ emission, whereas \lh\ retains absorption at this line and
cannot be securely assigned to luminosity class V or III at BOSS resolution. For \sk,
\niv\ emission and weak \niii\ establish the very-early-O regime, but the
uncertain ratio and absence of significant \nv\ motivate O2--3 rather than an
unqualified O2 subtype.
\par

Even if some of the targets are multiple, a luminous very-early-O component is
still required. Reproducible He~I in \wsi\ and \sk\ may reflect cooler companions
or fibre contamination, and the \sk\ radial-velocity trend is suggestive but
not significant. Such effects can modify line depths and inferred radius
scales, but do not naturally reproduce the high-ionisation morphology and
optical luminosity with an ordinary intermediate-mass stripped star as the
dominant component. Higher spectral resolution, higher signal-to-noise ratio (S/N), and denser time sampling are
required to establish multiplicity and quantitative atmosphere parameters.
\par

\section{Conclusions}
\label{sec:conclusions}

SDSS-V/BOSS spectroscopy establishes \wsi\ and \lh\ as new members of the LMC
very-early-O population and identifies \sk\ as an additional O2--3~V--III
source. We classify the three objects as O2~If$^\ast$, O2~V--III, and
O2--3~V--III, respectively. Independently, the Gaia CMD and OSTAR2002 scaling
place them in the luminous massive-star regime, several magnitudes above the
observed intermediate-mass stripped-star locus, consistent with their
identification as massive stars.
\par

The earlier classifications also show why such rare stars can remain
unrecognised: \wsi\ was obscured by its infrared-selected associations, the
MUSE classification of \lh\ lacked the decisive blue nitrogen region, and no
detailed blue-optical classification was available for \sk. Complete coverage
of \niv~$\lambda4058$, \nv~$\lambda\lambda4604,4620$, and
\niii~$\lambda\lambda4634$--4642 is therefore essential for recovering the
hottest normal O stars. SDSS-V/MWM is well suited to a systematic Magellanic-Cloud search;
higher-resolution follow-up will refine wind
morphology and multiplicity.
\par

\bibliographystyle{aa}
\bibliography{references_corrected_review_CMD_SED,references_Gotberg2023_addition}

\begin{appendix}
\nolinenumbers

\section{SDSS-V/BOSS observations and visit-resolved analysis}
\label{app:boss}

\FloatBarrier
For \wsi, four visit spectra
(\texttt{CATALOGID}=63050395787470304) comprise six individual exposures with
a total nominal exposure time of 5400~s. Three visits are available for \lh\
(\texttt{CATALOGID}=63050395782984402), with a total nominal exposure time of
2700~s. For \sk\ (\texttt{CATALOGID}=63050395785568943;
\texttt{SDSS\_ID}=91419225; Gaia Data Release 3 (DR3)
\texttt{source\_id}=4658095755697542528), the visits were obtained at
modified Julian dates (MJDs) 60212, 60213, and 60235 and comprised four nominal
900-s exposures.
\par

For the display combinations, each visit first received a broad 401-pixel
multiplicative response correction before inverse-variance combination on the
native logarithmic wavelength grid. Candidate artefacts were rejected only
when the other visits agreed and the discrepancy was
confined to an isolated run of one or two pixels; longer discrepant runs were
retained and reported rather than clipped. Recurrent stellar, wind, and
nebular features were therefore preserved. The BOSS wavelength scale is
expressed in vacuum, and vacuum rest wavelengths were used throughout the
radial-velocity analysis. Quantitative EW and velocity measurements were made
from the individual visits, not from the display combination.
\par

For the common stellar velocity, the available photospheric
He~II~$\lambda\lambda4200,4542,5412,6683,6891,7178$ absorption lines were
used where measurable. \heii~$\lambda4686$ was excluded because of its wind
sensitivity. Nitrogen-line velocities were used only as consistency checks,
and the blended \niii~$\lambda\lambda4634$--4642 complex was excluded because
its centroid is unstable at BOSS resolution.

\begin{table}[t]
\caption{Stellar and nebular velocities.}
\label{tab:velocities}
\centering
\begin{tabular}{lccc}
\hline\hline
Source & $v_{\rm He\,II}$ & $v_{\rm neb}$ & $\Delta v$ \\
       & (\kms) & (\kms) & (\kms) \\
\hline
\wsi & $295\pm17$ & $280.8\pm4.3$ & $+14\pm18$ \\
\lh  & $240\pm7$  & $251.3\pm4.4$ & $-11\pm8$ \\
\sk  & $277\pm7$  & $281\pm19$    & $-5\pm20$ \\
\hline
\end{tabular}
\end{table}

For \sk, He~II~$\lambda4200$, $\lambda4542$, and $\lambda5412$ were fitted
simultaneously in each visit with a common velocity shift and independent local
continua, depths, and widths. Table~\ref{tab:sk_visits} gives the resulting
visit velocities.

\begin{table}[t]
\caption{Visit-resolved radial velocities of \sk.}
\label{tab:sk_visits}
\centering
\begin{tabular}{lc}
\hline\hline
MJD & $v_{\rm He\,II}$ (\kms) \\
\hline
60212 & $299.7\pm13.8$ \\
60213 & $288.3\pm15.3$ \\
60235 & $260.5\pm11.2$ \\
\hline
\end{tabular}
\end{table}

\FloatBarrier
\section{Preparation of the LMC O2 comparison spectra}
\label{app:templates}

\FloatBarrier
BI~237 and Brey~77 were observed with the ultraviolet--blue (UVB) arm of the
Very Large Telescope (VLT)/X-shooter under European Southern Observatory (ESO)
programme 106.211Z.001. Both products were obtained with the
$0.8\times11$~arcsec slit and have a reference resolving power of $R=6655$.
The BI~237 and Brey~77 archive products are
\texttt{ADP.2021-03-08T12:59:03.267.fits} and
\texttt{ADP.2021-01-11T12:13:26.255.fits}, respectively.
\par

The 30~Dor~016 template was constructed from stacked VLT/Fibre Large Array
Multi Element Spectrograph (FLAMES)--GIRAFFE LR02 and LR03 products obtained
under programme 182.D-0222(A). The LR02 setting covers
3959.99--4570.99~\AA\ at $R=6300$, whereas LR03 covers
4498.98--5077.98~\AA\ at $R=7500$. The two settings overlap between
approximately 4499 and 4571~\AA, allowing their relative continuum placement
to be verified before the two spectral segments were joined.
\par

Each template was shifted to its stellar rest frame and pseudo-continuum
normalised over the blue-optical interval. Invalid pixels and isolated
reduction artefacts were masked, while recurrent stellar, wind, and nebular
features were retained. Because the archival templates have substantially
higher spectral resolution than the BOSS data, they were convolved to the
common wavelength-dependent instrumental width adopted for Fig.~1. The required
Gaussian full width at half maximum (FWHM) of the kernel was calculated at each
wavelength from
\[
 {\rm FWHM}_{\rm ker}(\lambda)=
 \left[
 {\rm FWHM}_{\rm common}^{2}(\lambda)-
 {\rm FWHM}_{\rm temp}^{2}(\lambda)
 \right]^{1/2},
\]
where ${\rm FWHM}_{\rm common}(\lambda)$ is defined by the largest effective
BOSS \texttt{WRESL} among the three science spectra. The corresponding common
resolving power increases from approximately $R=990$ at 4000~\AA\ to $R=1322$
at 5060~\AA. No rotational broadening or adjustment of individual line
strengths or profiles was applied. This homogenisation was used only for the
morphological display in Fig.~1.
\par

\FloatBarrier
\section{Statistical interpretation of the nitrogen diagnostic}
\label{app:nitrogen}

\FloatBarrier
\begin{table}[H]
\caption{Inverse-variance-weighted mean equivalent widths of the principal
classification diagnostics measured in the individual SDSS-V/BOSS visits.}
\label{tab:key_lines_three}
\centering
\resizebox{\columnwidth}{!}{
\begin{tabular}{lccc}
\hline\hline
Feature & \wsi & \lh & \sk \\
        & EW (\AA) & EW (\AA) & EW (\AA) \\
\hline
\niv~$\lambda4058$  & $-0.75\pm0.18$ & $-0.82\pm0.15$ & $-0.38\pm0.20$ \\
\niii~$\lambda4640$ & $-0.17\pm0.11$ & $-0.27\pm0.14$ & $-0.17\pm0.15$ \\
\hei~$\lambda4471$  & $+0.31\pm0.10$ & $+0.23\pm0.12$ & $+0.22\pm0.09$ \\
\heii~$\lambda4542$ & $+0.67\pm0.15$ & $+0.79\pm0.11$ & $+0.67\pm0.13$ \\
\nv~$\lambda4604$   & $+0.20\pm0.07$ & $+0.31\pm0.08$ & $+0.04\pm0.08$ \\
\nv~$\lambda4620$   & $+0.27\pm0.08$ & $+0.24\pm0.08$ & $-0.01\pm0.13$ \\
\heii~$\lambda4686$ & $-1.61\pm0.37$ & $+0.80\pm0.11$ & $+0.39\pm0.08$ \\
H$\beta$            & $+0.61\pm0.23$ & $+1.90\pm0.17$ & $+2.32\pm0.72$ \\
\hline
\end{tabular}
}
\tablefoot{Positive EWs denote absorption, whereas negative values denote emission. Each
visit was measured independently after rest-frame correction. For \wsi\ and
\lh, H$\beta$ is affected by narrow nebular-emission infilling, as indicated
independently by [O~{\sc iii}]~$\lambda\lambda4959,5007$; the quoted H$\beta$
EWs should therefore not be interpreted as purely stellar line strengths.}
\end{table}

For \wsi\ and \lh, both the measured \niv/\niii\ ratios and the detection of
\nv~$\lambda\lambda4604,4620$ support an O2 classification. Their weighted logarithmic ratios are $0.64\pm0.29$ and $0.48\pm0.23$,
respectively. For \sk, both the numerator and denominator are weak and the
weighted ratio, $0.34\pm0.44$, is correspondingly imprecise; neither \nv\
component is significantly detected.
\par

For \sk, the classification therefore rests mainly on the very-early-O morphology:
\niv\ emission, weak \niii, strong He~II absorption, and the close similarity to
BI~237, rather than on the exact central value of the formal \niv/\niii\ ratio.
The lack of a significant \nv\ detection should not, however, be treated as proof
of intrinsic absence, because shallow or rotationally broadened \nv\ absorption
can be difficult to recover at modest S/N \citep{Walborn2002}. Given these
limitations, we adopt the conservative O2--3 subtype range for \sk\ instead of an
unqualified O2 classification.
\par

\FloatBarrier
\section{Broad-band SED modelling and Gaia colour--magnitude diagram}
\label{app:sed}

\FloatBarrier
\subsection{OSTAR2002 SED fits and infrared diagnostics}

The broad-band analysis uses the OSTAR2002 non-local thermodynamic equilibrium
(NLTE), line-blanketed atmosphere grid \citep{LanzHubeny2003}. We adopt the half-solar grid at $\log g=4.0$ and
$d=49.6$~kpc. Because the available optical and near-infrared photometry samples the
Rayleigh--Jeans side of such hot SEDs, it is not used to determine
$T_{\rm eff}$. Instead, the five exact OSTAR2002 nodes at 45.0, 47.5, 50.0,
52.5, and 55.0~kK are sampled with equal probability. At each Monte-Carlo
draw, catalogue uncertainties and an additional 2\% flux floor are propagated
while $A_V$ and $R_\star$ are fitted jointly; $\log L/L_\odot$ and the Gaia
broad-band extinctions are then derived for that draw.
\par

We use the audited no-$U$ photometric variants. For \wsi\ and \sk\ the nominal
optical constraints are the Massey $V$, $B-V$, and $V-R$ measurements
\citep{Massey2002}, together with Two Micron All Sky Survey (2MASS) $JHK_s$
\citep{Skrutskie2006}; for \lh\ we use Magellanic Clouds Photometric Survey
(MCPS) $BVI$ photometry \citep{Zaritsky2004} and 2MASS. The same
selection procedure is applied to the early-O comparison stars: BI~237 uses
MCPS $BV$ and 2MASS $JHK_s$ (the MCPS $I$ datum is excluded after the
leave-one-out consistency test identified it as the dominant discrepant
measurement), whereas 30~Dor~016 uses Massey $V$, $B-V$, and $V-R$ together
with 2MASS $JHK_s$. For \wsi, an MCPS solution is propagated separately as a
catalogue-systematic cross-check rather than merged into the nominal posterior.
The $U$ band is excluded throughout. Extinction within the modelled
optical--near-infrared domain follows the LMC-average curve of \citet{Gordon2003}.
The Gaia $G$, $G_{\rm BP}$, and $G_{\rm RP}$ extinctions are obtained by
integrating the extinguished OSTAR2002 spectra through the corresponding
passbands.
\par

The nominal optical/near-infrared observables used in these fits are listed in Table~\ref{tab:sed_input_photometry}.
\par

\begin{table*}[t]
\caption{Optical and near-infrared photometry adopted in the nominal OSTAR2002 fits.}
\label{tab:sed_input_photometry}
\centering
\small
\begin{tabular}{p{0.12\textwidth}p{0.40\textwidth}p{0.34\textwidth}p{0.08\textwidth}}
\hline\hline
Source & Optical catalogue and native observables & 2MASS $JHK_s$ & Fit status \\
\hline
\wsi & Massey: $V=14.200\pm0.010$, $B-V=-0.040\pm0.010$, $V-R=+0.080\pm0.030$
     & $J=14.314\pm0.036$, $H=14.442\pm0.063$, $K_s=14.476\pm0.094$ & nominal \\
\lh & MCPS: $B=13.443\pm0.029$, $V=13.404\pm0.041$, $I=13.570\pm0.062$
     & $J=13.717\pm0.062$, $H=13.757\pm0.051$, $K_s=13.895\pm0.074$ & nominal \\
\sk & Massey: $V=12.980\pm0.010$, $B-V=-0.200\pm0.010$, $V-R=-0.070\pm0.020$
     & $J=13.456\pm0.030$, $H=13.566\pm0.041$, $K_s=13.617\pm0.062$ & nominal \\
BI~237 & MCPS: $B=13.790\pm0.023$, $V=13.830\pm0.043$
     & $J=13.950\pm0.027$, $H=14.011\pm0.043$, $K_s=14.039\pm0.072$ & nominal \\
30~Dor~016 & Massey: $V=13.490\pm0.010$, $B-V=+0.040\pm0.010$, $V-R=+0.060\pm0.030$
     & $J=13.384\pm0.041$, $H=13.345\pm0.061$, $K_s=13.356\pm0.057$ & nominal \\
\hline
\end{tabular}
\tablefoot{Magnitudes and colours are listed in the native catalogue observables actually fitted; in particular, Massey $B$ and $R$ magnitudes are not reconstructed or treated as independent data. The $U$ band is excluded for all objects. For BI~237, the MCPS $I=13.948\pm0.034$ measurement was examined but excluded from the nominal solution after the leave-one-out test identified it as the dominant discrepant datum. For \wsi, the alternative MCPS measurements $B=14.237\pm0.024$, $V=14.363\pm0.029$, and $I=14.181\pm0.026$ (with the same 2MASS counterpart) are propagated only as a catalogue-systematic cross-check and are not merged into the nominal posterior. A 2\% flux floor is added in quadrature during fitting.}
\end{table*}

\begin{figure*}[t]
\centering
\includegraphics[width=\textwidth]{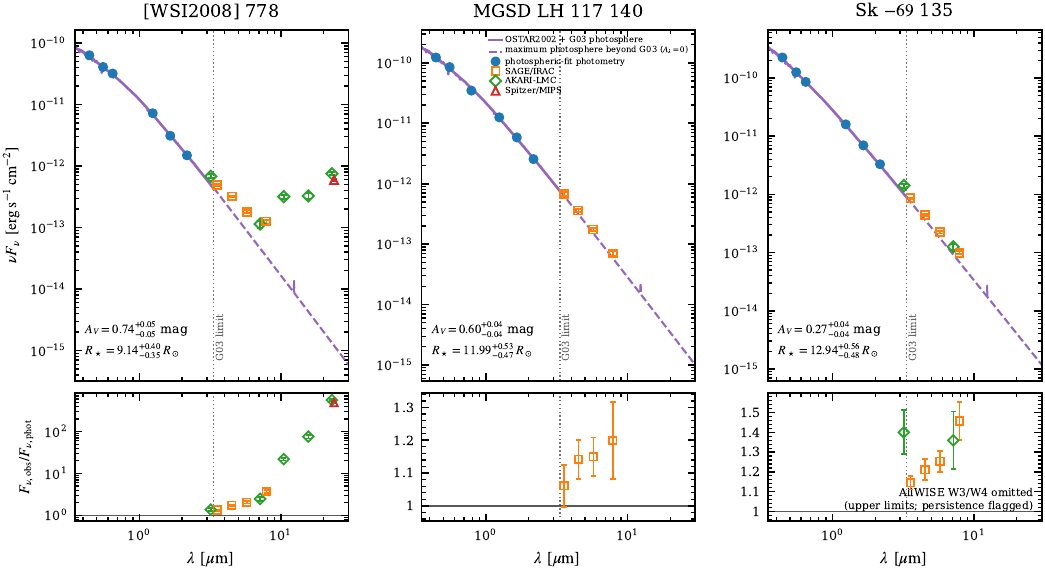}
\caption{Broad-band SEDs of \wsi\ (left), \lh\ (middle), and \sk\ (right).
Upper panels show $\nu F_\nu$, and the lower panels show the observed-to-photospheric
flux ratio. Within the formal Gordon et al. (2003) domain, the solid curve is the median
OSTAR2002 photosphere reddened with their LMC-average curve, and the shaded
interval represents the 16th--84th
percentiles of the Monte-Carlo model. The vertical dotted line marks
$3.33\,\mu$m. At longer wavelengths, the dashed curve is the conservative
maximum photosphere obtained with $A_\lambda=0$; no MIR extinction law is
extrapolated. Filled circles are the optical/near-infrared constraints used in the
photospheric fit, while open symbols show the selected SAGE/IRAC, AKARI-LMC,
and, where available, Spitzer/MIPS (Multiband Imaging Photometer for Spitzer)
measurements. \wsi\ displays a strong
rising MIR excess; \lh\ remains close to the photosphere; \sk\ shows a modest
$3$--$8\,\mu$m excess. AllWISE $W3$ and $W4$ for \sk\ are not used because
the catalogue reports upper-limit photometric-quality flags and persistence
contamination in both bands.}
\label{fig:sed_three_sources}
\end{figure*}

Figure~\ref{fig:sed_three_sources} compares the photospheric predictions with
selected mid-infrared measurements. Spitzer Surveying the Agents of Galaxy
Evolution (SAGE) photometry from the Infrared Array Camera (IRAC) is taken from
the LMC survey \citep{Meixner2006}, AKARI points from the LMC point-source
catalogue (VizieR J/AJ/144/179/lmccat01; \citealt{Kato2012}), and the
24-$\mu$m point of \wsi\ from the SAGE massive-YSO compilation
\citep{Whitney2008}. The adopted
Gordon et al. extinction curve is used only over its formal wavelength domain,
ending at $\lambda=3.33\,\mu$m. Beyond this boundary we do not extrapolate an
extinction law: the dashed curve is the maximum stellar photosphere obtained
by setting $A_\lambda=0$. Thus an observed point above that curve represents a
conservative excess relative even to the largest allowed stellar continuum.
No dust component is fitted.
\par

The MIR measurements shown in Fig.~\ref{fig:sed_three_sources} are listed in Table~\ref{tab:sed_mir_photometry}.
\par

\begin{table*}[t]
\caption{Selected mid-infrared diagnostic photometry displayed in Fig.~\ref{fig:sed_three_sources}.}
\label{tab:sed_mir_photometry}
\centering
\small
\begin{tabular}{p{0.12\textwidth}p{0.35\textwidth}p{0.38\textwidth}p{0.10\textwidth}}
\hline\hline
Source & SAGE/IRAC fluxes (mJy) & AKARI-LMC fluxes (mJy) & Other \\
\hline
\wsi & 3.6: $0.590\pm0.022$; 4.5: $0.480\pm0.017$; 5.8: $0.346\pm0.020$; 8.0: $0.330\pm0.024$
 & N3: $0.725\pm0.055$; S7: $0.271\pm0.035$; S11: $1.110\pm0.080$; L15: $1.700\pm0.180$; L24: $5.740\pm0.330$
 & MIPS~24: $4.690\pm0.210$ \\
\lh & 3.6: $0.802\pm0.048$; 4.5: $0.540\pm0.028$; 5.8: $0.334\pm0.017$; 8.0: $0.184\pm0.018$
 & -- & -- \\
\sk & 3.6: $1.010\pm0.030$; 4.5: $0.668\pm0.029$; 5.8: $0.425\pm0.018$; 8.0: $0.261\pm0.017$
 & N3: $1.510\pm0.120$; S7: $0.298\pm0.032$ & -- \\
\hline
\end{tabular}
\tablefoot{These measurements are diagnostic only and are excluded from the optical--near-infrared photospheric fit. SAGE/IRAC values are taken from the selected \texttt{II/305/catalog} entries, AKARI-LMC values from \texttt{J/AJ/144/179/lmccat01}, and the \wsi\ MIPS point from \texttt{J/AJ/136/18/table12}. When more than one entry from the selected catalogue product exists for a band, the plotted flux is the median and its adopted uncertainty is the larger of the median quoted uncertainty and the row-to-row scatter. AllWISE $W3/W4$ measurements of \sk\ are not listed because they are upper-limit/persistence-affected entries and were not used as detections.}
\end{table*}

The posterior median $(A_V,R_\star/R_\odot)$ values are
$(0.74,9.14)$, $(0.60,11.99)$, and $(0.27,12.94)$ for \wsi, \lh, and \sk,
respectively; these radii are conditional on the adopted 45--55~kK atmosphere
range and are scaling checks rather than temperature determinations.
\par

The qualitative result is not sensitive to the remaining temperature--radius
covariance. For \wsi, the excess grows strongly from the IRAC bands towards the
AKARI and MIPS measurements, which can explain its previous infrared-selected classifications without implying
that the optical spectrum itself is that of a YSO or post-AGB star. The excess may arise from circumstellar or local
environmental dust, unresolved multiplicity, blending, or a combination of
these effects. \lh\ remains essentially photospheric over the available MIR
coverage. For \sk, independent IRAC and AKARI measurements indicate only a
mild systematic departure at $3$--$8\,\mu$m; the previously apparent very
large long-wavelength Wide-field Infrared Survey Explorer (WISE) excess is not
retained because $W3$ and $W4$ are
upper limits affected by persistence flags.
\par

\subsection{Gaia colour--magnitude diagram}

The same Monte-Carlo realisations provide $A_G$, $A_{\rm BP}$, and $A_{\rm RP}$,
which are combined with Gaia DR3 photometry \citep{GaiaDR3} to form
\[
 (G_{\rm BP}-G_{\rm RP})_0=(G_{\rm BP}-G_{\rm RP})-(A_{\rm BP}-A_{\rm RP})
\]
and
\[
 M_{G,0}=G-A_G-5\log(d/10\,{\rm pc}).
\]
For the five O-star objects, the adopted CMD coordinates are the posterior
medians of these quantities, and the 16th--84th percentile intervals are shown
as asymmetric error bars in Fig.~\ref{fig:gaia_cmd}. BI~237 and 30~Dor~016 are
processed through the same OSTAR2002/extinction framework and provide empirical
very-early-O comparisons. The observed Gaia photometry and the final intrinsic
CMD coordinates are tabulated in Table~\ref{tab:cmd_photometry}.
\par

\begin{table*}[t]
\caption{Gaia DR3 photometry and extinction-corrected CMD coordinates.}
\label{tab:cmd_photometry}
\centering
\small
\begin{tabular}{lrrrrrr}
\hline\hline
Source & Gaia DR3 source\_id & $G$ & $G_{\rm BP}$ & $G_{\rm RP}$ & $(G_{\rm BP}-G_{\rm RP})_0$ & $M_{G,0}$ \\
\hline
\wsi & 4658432721684644480 & 14.256787 & 14.193605 & 14.279823 & $-0.509^{+0.033}_{-0.034}$ & $-4.969^{+0.052}_{-0.050}$ \\
\lh & 4657383027359753728 & 13.519092 & 13.427448 & 13.600076 & $-0.514^{+0.025}_{-0.025}$ & $-5.563^{+0.050}_{-0.050}$ \\
\sk & 4658095755697542528 & 12.970034 & 12.854306 & 13.174096 & $-0.478^{+0.023}_{-0.023}$ & $-5.788^{+0.047}_{-0.047}$ \\
BI~237 & 4660109236409051392 & 13.825301 & 13.778407 & 13.879809 & $-0.524^{+0.021}_{-0.021}$ & $-5.400^{+0.043}_{-0.043}$ \\
30~Dor~016 & 4657690620070706432 & 13.479229 & 13.482241 & 13.430018 & $-0.562^{+0.024}_{-0.024}$ & $-6.081^{+0.049}_{-0.049}$ \\
\hline
G\"otberg Star~1 & -- & -- & -- & -- & $\simeq-0.560$ & $\simeq-1.80$ \\
\hline
\end{tabular}
\tablefoot{The Gaia magnitudes are the observed DR3 mean photometry. For the five O-star objects, the intrinsic colour and absolute magnitude are posterior medians; superscripts and subscripts give the 84th-minus-median and median-minus-16th-percentile intervals, respectively, from the homogeneous OSTAR2002/extinction Monte-Carlo propagation. The G\"otberg Star~1 values were obtained using the approximate Gaia-equivalent external transformation described in the text and do not carry a homogeneous Monte-Carlo interval.}
\end{table*}

Figure~\ref{fig:gaia_cmd} compares the resulting positions with PARSEC v1.2S
$Z=0.006$ isochrones at 0.05, 0.1, 0.2, 0.5, 1, and 2~Myr
\citep{Bressan2012}. At very-early-O temperatures,
$G_{\rm BP}-G_{\rm RP}$ is only weakly sensitive to $T_{\rm eff}$, so small
horizontal reversals among adjacent young isochrones are not interpreted as a
physical temperature inversion. Initial-mass labels on the 1 and 2~Myr curves are shown only as guides. The reddening vector corresponds to $A_V=0.2$~mag
and is derived with the same OSTAR2002 plus LMC-average extinction framework
used for the homogeneous CMD corrections. The CMD is used to establish the
luminosity regime rather than to infer precise ages or masses.
\par

For comparison with binary-stripped objects, we considered the observational
sample of intermediate-mass stripped helium stars analysed with CMFGEN by
\citet{Gotberg2023}. Stars~1--8, for which the stripped star dominates the
optical light, span approximately $T_{\rm eff}=57$--90~kK and
$\log L/L_\odot=3.5$--5.1, so stripped stars can be as hot as, or hotter than, the earliest O stars. Applying the same hot-star synthetic
Gaia-colour mapping used as an external diagnostic places this hot subsample
at approximately $-0.56\lesssim(G_{\rm BP}-G_{\rm RP})_0\lesssim-0.51$;
thus colour alone does not distinguish a hot stripped star from an early-O
star. Their optical luminosities are nevertheless much lower. Using the
published AB photometry, extinctions, and Magellanic-Cloud distances gives
approximately $-1.7\lesssim M_{V,\rm AB}\lesssim+0.8$ for Stars~1--8.
Star~1, the optically brightest member of this hot stripped-star sample, is
therefore shown in Fig.~\ref{fig:gaia_cmd} as a single representative point.
Its plotted Gaia coordinate is an approximate hot-star transformation of the
published solution and is not part of the OSTAR2002 Monte-Carlo analysis.
\par

The main distinction is therefore optical luminosity rather than intrinsic
colour. The three science targets, with
$M_{G,0}=-4.97$, $-5.56$, and $-5.79$~mag, lie several magnitudes above the
observed intermediate-mass stripped-star regime and occupy the same luminous
blue domain as BI~237 and 30~Dor~016. Their CMD positions therefore strongly
disfavour an interpretation in which the optically dominant components are
ordinary intermediate-mass stripped helium stars, consistent with the small
radii expected for binary-stripped products \citep{Gotberg2018,Gotberg2023}.
This comparison does not exclude binary evolution: unresolved companions, past
interaction, and more massive binary-evolution products remain possible. We
therefore use the CMD to establish the luminosity regime, not to assign precise
ages, initial masses, or single-star evolutionary histories.
\par

\FloatBarrier
\section{Acknowledgements}\label{app:acknowledgements}

\FloatBarrier
This research is based in part on observations collected at the European
Southern Observatory under ESO programmes 106.211Z.001 and 182.D-0222(A), and
on data obtained from the ESO Science Archive Facility. The archival spectra
used in this work comprise the VLT/X-shooter observations of BI~237 and
Brey~77 obtained under programme 106.211Z.001 and the
VLT/FLAMES--GIRAFFE observations of 30~Dor~016 obtained under programme
182.D-0222(A).

Funding for the Sloan Digital Sky Survey V has been provided by the Alfred P.
Sloan Foundation, the Heising-Simons Foundation, the National Science
Foundation, and the Participating Institutions. SDSS acknowledges support and
resources from the Center for High-Performance Computing at the University of
Utah. SDSS telescopes are located at Apache Point Observatory, funded by the
Astrophysical Research Consortium and operated by New Mexico State University,
and at Las Campanas Observatory, operated by the Carnegie Institution for
Science. CRZ acknowledges financial support from the UNAM-DGAPA-PAPIIT
IN107226. J.G.F.-T. gratefully acknowledges support from ANID Fondecyt Regular
No. 1260371. The SDSS web site is \url{www.sdss.org}.

SDSS is managed by the Astrophysical Research Consortium for the Participating
Institutions of the SDSS Collaboration, including the Carnegie Institution for
Science, Chilean National Time Allocation Committee (CNTAC) ratified
researchers, Caltech, the Gotham Participation Group, Harvard University,
Heidelberg University, The Flatiron Institute, The Johns Hopkins University,
L'Ecole polytechnique f\'{e}d\'{e}rale de Lausanne (EPFL),
Leibniz-Institut f\"{u}r Astrophysik Potsdam (AIP),
Max-Planck-Institut f\"{u}r Astronomie (MPIA Heidelberg),
Max-Planck-Institut f\"{u}r Extraterrestrische Physik (MPE), Nanjing
University, National Astronomical Observatories of China (NAOC), New Mexico
State University, The Ohio State University, Pennsylvania State University,
Smithsonian Astrophysical Observatory, Space Telescope Science Institute
(STScI), the Stellar Astrophysics Participation Group, Universidad Nacional
Aut\'{o}noma de M\'{e}xico, University of Arizona, University of Colorado
Boulder, University of Illinois at Urbana-Champaign, University of Toronto,
University of Utah, University of Virginia, Yale University, and Yunnan
University.

\FloatBarrier
\end{appendix}

\end{document}